\documentclass[useAMS,usenatbib]{mn2e}
\usepackage{graphicx}
\usepackage{amssymb}

\title[Dark Matter in Cluster Dwarf Spheroidals]{Hubble Space Telescope survey of the Perseus Cluster -\\I: The structure and dark matter content of cluster dwarf spheroidals}
\author[S. J. Penny et al.]{Samantha~J.~Penny$^1$, Christopher~J.~Conselice$^1$, Sven~De~Rijcke$^2$ and Enrico~V.~Held$^3$ \footnotemark[0]\\ 
$^1$School of Physics \& Astronomy, University of Nottingham, Nottingham, NG7 2RD, United Kingdom\\
$^2$Sterrenkundig Observatorium, Universiteit Gent, Krijgslaan 281, S9, B-9000, Gent, Belgium\\
$^3$Osservatorio Astronomico di Padova, INAF, Vicolo Osservatorio 5, I-35122 Padova, Italy}

\begin{document}

\maketitle

\begin{abstract}
We present the results of a Hubble Space Telescope ($HST$) Advanced Camera for Surveys (ACS) study of dwarf galaxies in the core of the rich nearby Perseus Cluster, down to M$_{V}~=~-12$. We identify 29 dwarfs as cluster members, 17 of which are previously unstudied. All the dwarfs we examine are remarkably smooth in appearance, and lack internal features. This smoothness is quantified by parametric fitting of the two-dimensional light distribution, and through non-parametric structural parameters. The dwarfs in this study are found to be very smooth, and have very low asymmetry and clumpiness values ($<\rm{A}>~=~0.03~\pm~0.04$, $<\rm{S}>~=~0.02~\pm~0.09$) throughout their entire structure, showing no evidence for internal features or star formation that could be the result of tidal processes or star formation induced by the cluster environment. Based on these observations, and the sizes of
these dwarfs, we argue that some of the dwarfs in our sample must have a large dark matter content to prevent disruption by the cluster potential. We derive a new method, independent of kinematics, for measuring the dark matter content of dEs, based on the radius of the dwarf, the projected distance of the dwarf from the cluster centre, and the total mass of the cluster interior to it. We use this method to determine the mass-to-light ratios for 25 of the dwarfs in our sample, the others being too close to bright stars to accurately determine their radii. At their current positions in the cluster, 12 out of 25 of the dwarfs in our sample require dark matter to remain stable against the cluster potential. These 12 dwarfs are all fainter than M$_{V} = -14$, and nearly all dEs with M$_{V} < -13$ require dark matter to survive. We also consider the mass that the dwarfs would require to survive the tidal forces at the centre of the cluster, at a pericentric distance of 35 kpc from the cluster centre. We find that at this distance,  the mass-to-light ratios of these dwarfs are comparable to those of the Local Group dSphs, ranging between M$_{\odot}$/L$_{\odot}$ $\approx$ 1 and 120. 
\end{abstract}


\section{Introduction}

Dwarf elliptical (dE) galaxies (M$_{\rm{B}} > -18$), and the fainter dwarf spheroidal (dSph) galaxies (M$_{\rm{B}} > -14$), are the most numerous galaxy type in the Universe, and are thus fundamental for understanding galaxy formation.  Dwarf galaxies are furthermore believed to be among the most dark matter dominated objects, with mass-to-light ratios up to a few hundred for Local Group dwarf spheroidals (e.g. \citealt*{lokas05,strigari07}).  A deeper understanding of dark matter and its role in galaxy formation can thus potentially be obtained through observing these lowest mass galaxies.

Measuring the dark matter content of dwarf galaxies has almost exclusively been carried out using kinematic measures of individual stars in Local Group systems (e.g., \citealt*{mateo08}). Thus, one method for determining the dark matter content, and therefore mass-to-light ratios, of cluster dwarf ellipticals is to measure their internal velocity dispersions (assuming the dwarfs are in virial equilibrium). However, due to the large distances to galaxy clusters it is impossible to resolve the stellar populations of cluster dEs/dSphs, and thus kinematics cannot be easily used to find the dark matter content of such galaxies.  

Using kinematics to measure the dark matter content of a dwarf furthermore has additional drawbacks. Firstly, unresolved binary stars can increase the measured velocity dispersion, resulting in larger measured M/L ratios for the dwarf \citep{KlessenKroupa,sven02}. Measuring the mass-to-light ratio of a dwarf using its measured velocity dispersion also assumes that the dwarf is in virial equilibrium. However, a satellite galaxy may be significantly disturbed by tides (e.g. \citealt{KlessenKroupa}), which could result in the dwarf being sufficiently perturbed from virial equilibrium that the total mass will be incorrectly measured using kinematics. Therefore, independent methods for estimating the dark-to-light ratios for dwarfs are needed that do not involve the use of kinematics.

We derive in this paper a new method for determining the dark matter content of dwarf elliptical galaxies in dense cluster environments, without the use of kinematics. This method is based on the smoothness and symmetry of light profiles, the position of the dwarf in the cluster,  as well as the size of the dwarf. We measure the mass-to-light ratios for dwarfs in the Perseus cluster by finding the minimum total mass that each galaxy must have to prevent it from being disrupted by the cluster potential at its current projected cluster centre distance in the cluster. We also test how much dark matter would be required for dwarfs to survive in the centre of the Perseus Cluster, at a pericentric distance of 35 kpc from the cluster centered galaxy NGC 1275. 

We  utilise high resolution \textit{Hubble Space Telescope (HST)} Advanced Camera for Surveys (ACS) imaging of the Perseus Cluster in the F555W and F814W bands to identify 29 dEs/dSph systems in the core of Perseus. Seventeen of these dwarfs were identified by \citet*{cons03}, with the remaining twelve galaxies new identifications. Six of these dEs are spectroscopically confirmed members of the Perseus Cluster \citep{me08}, and the remainder are easily differentiated from background galaxies. The dwarfs we identify are smooth in appearance, lacking internal features, suggesting they have not been tidally disturbed by the cluster potential. We infer from this smoothness that the dwarfs  must have a large dark matter content to prevent tidal disruption, and calculate the mass-to-light ratios for the dwarfs in our sample using the new method presented in this paper. We find that at their current positions in the cluster, 12 of the dwarfs require dark matter. We also consider the smallest distance from the cluster centre at which a dwarf can survive the cluster potential, which we take to be 35 kpc. At this distance from the cluster centre, the dwarfs have mass-to-light ratios comparable to those of the Milky Way dSphs, ranging between M$_{\odot}$/L$_{\odot}$ $\approx$ 1 and 120.

This paper is organized as follows. In Section 2, we discuss our observations. Section 3 covers the results of these observations, including the criteria for cluster membership discussed in Section 3.1, structural parameters in Section 3.2, and surface photometry in Section 3.3. The implications for measuring dark matter, based on the smoothness of our dwarfs is covered in Section 4, along with our new method to determine the dark matter content of the dwarfs. The dark matter content of the dwarfs is presented in Section 5, with a discussion of these results in Section 6. We summarize our results in Section 7. Throughout this paper we will refer to both cluster dEs and cluster dSphs as dEs.

\section{Observations}

We examine the dwarf galaxy population in the Perseus Cluster (Abell 426) which is one of the richest nearby galaxy clusters, with a redshift $v$ = 5366 km s$^{-1}$ \citep{Stublerood99}, and at a distance $D$ = 71 Mpc. Due to its low galactic latitude ($b$ $\approx$ $-13^{\circ}$) it has not been studied in as much detail as other nearby clusters such as Fornax, Virgo and Coma.  The Perseus cluster is host to the unusual and X-ray bright central galaxy NGC~1275 \citep*{conselice01b}, and has been the subject of dwarf galaxies studies in the past (\citealt*{conselice02}, \citealt{cons03}).

As part of a general study of these dwarf galaxy population we obtained high resolution \textit{Hubble Space Telescope (HST)} Advanced Camera for Surveys (ACS) WFC imaging in the F555W and F814W bands of five fields in the Perseus Cluster core. Exposure times are one orbit each in the F555W and F814W bands. The scale for these images is 0.05'' pixel$^{-1}$, with a field of view of 202'' $\times$ 202'', providing a total $HST$ survey area of $\sim$ 57 arcmin$^{2}$.

As our analysis involves fitting and measuring photometric features for our dwarfs we cleaned and remove from our ACS imaging foreground stars and background galaxies using the IRAF routine IMEDIT. The fluxes within a 1'' aperture were then measured, and converted to apparent F555W and F814W magnitudes using the zero-points of \citet{sirianni}. These magnitudes were then converted to the $UBVRI$ system using the transformations of \citet{sirianni}, allowing for the measurement of $(V-I)$ colours.   Total magnitudes for our dwarf sample are found using our own code, which fits an ellipse through a set of positions where a given surface brightness level is reached. Foreground stars and background stars were masked out and were not used in this fit. Total magnitudes are measured from the fitted model, and converted to $V$ and $I$ magnitudes using the same zero-points and transformations as for the central 1'' colours. 

\section{Observational Results}

\subsection{Cluster Membership}

The most reliable method for identifying cluster membership is by measuring radial velocities, although this is not easily done, or even possible, for low surface brightness objects, such as dwarf ellipticals. Previous to this paper confirmed cluster membership existed for six of the dwarfs based on spectroscopic observations \citep{me08}. All the galaxies we examine in this study are well resolved, thus if an object has the same morphology as the spectroscopically confirmed cluster members, we assume that it is also a cluster member. Deep, high resolution $HST$ ACS imaging reveals detail not seen in ground-based observations, making the identification of background members easier, as such galaxies are often spirals. We identify by eye all galaxies with similar morphologies to the spectroscopically confirmed cluster members. Using this method, we identify a total of 29 dwarf ellipticals from our ACS imaging.

\begin{figure}
\includegraphics[width=84mm]{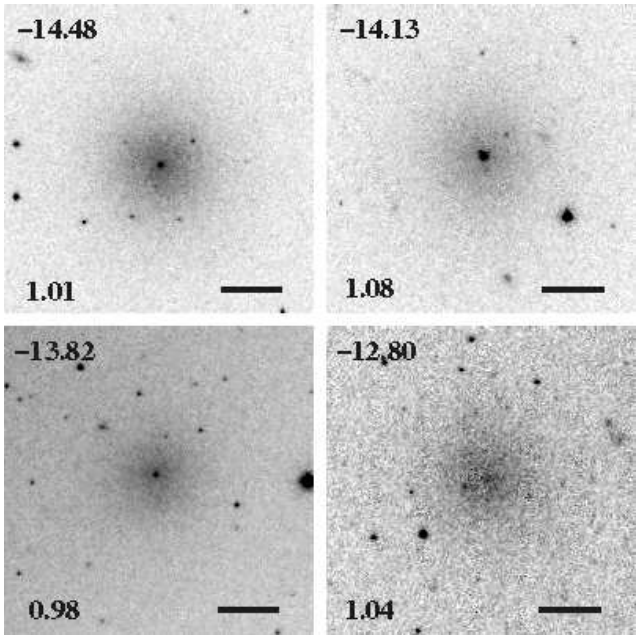}
\caption{Selection of Perseus dwarfs used in this study. The upper left number is the value of M$_{\rm{V}}$ for the dwarf, and the lower left number the ($V-I$) colour. The solid bar is 2'' in length.}
\end{figure}

Figure 1 shows a selection of four Perseus cluster dwarfs used in this study. The dwarfs are remarkably smooth and symmetric in appearance, without evidence for substructure which could indicate star formation, spiral arms or other internal features. They are also apparently spherical, suggesting that they have not been tidally disrupted by the dense cluster environment. Penny et al. (2008, in preparation) will include a more detailed description and catalogue of these dwarfs.

\subsection{CAS parameters} 

We use the concentration, asymmetry and clumpiness (CAS) parameters to determine the non-parametric structures of our dwarf sample \citep{cons03b}. We measure the CAS parameters using the inverted form of the Petrosian radius \citep{Pet76,Kron95} for each galaxy. Petrosian radii have been found to be the most consistent and robust radii to use in the derivation of structural parameters of galaxies \citep*{Bershady00,Conselice00}. We do not include any dwarfs near bright foreground stars in this part of the study, as the Petrosian radius could not be accurately measured in these cases.
 
The Petrosian radius for each galaxy is defined as:
\begin{equation}
 R_{Petr} = 1.5 \times r(\eta = 0.2),
\end{equation}
\noindent where $\eta$ is the ratio of the average surface brightness with a radius $r$ to the local surface brightness at $r$. The $\eta = 0.2$ radius was chosen as it contains more than 99\% of the light for galaxies with exponential profiles \citep{Bershady00}.

We compute the concentration index (C) for each galaxy, which is defined as the ratio of the radii containing 80$\%$ and 20$\%$ of a galaxy's light. The higher the value of C, the more concentrated the light of the galaxy is towards the centre, with dwarf ellipticals having an average value of 2.5 $\pm$ 0.3 \citep{cons03b}. The asymmetry indices (A) for the dwarfs in our sample are also computed. This index measures the deviation of the galaxies light from perfect 180$^{\circ}$ symmetry. The less symmetric the galaxy, the higher the asymmetry index. Asymmetric light distributions are produced by features such as star formation, galaxy interactions/mergers and dust lanes \citep{Conselice00}, none of which are expected to be seen in dwarf ellipticals. Therefore dEs are expected to have low asymmetries of A $\approx$ 0.

The third index we measure is the clumpiness index (S). This index describes how patchy the distribution of light within a galaxy is. By simply observing the dwarfs by eye, it is clear that they are smooth systems, with little to no clumpiness in their light distributions.   The clumpiness (S) index is the ratio of the amount of light contained in high frequency structures to the total amount of light in the galaxy. To measure S, the original effective resolution of the galaxy is reduced by smoothing the galaxy by a filter of width $\sigma = R_{Petr}/5$ to create a new image. The original image is then subtracted by the new, lower resolution, image to produce a residual map. The flux of the residual light is then summed, and this sum is divided by the sum of the flux in the original galaxy image to give the galaxies clumpiness (S). For the dwarf sample, this ratio should be near zero. This index is very sensitive to signal-to-noise, and we found it necessary to compute this index at half the Petrosian radius of each dwarf.

\begin{figure*}
\includegraphics[width=170mm]{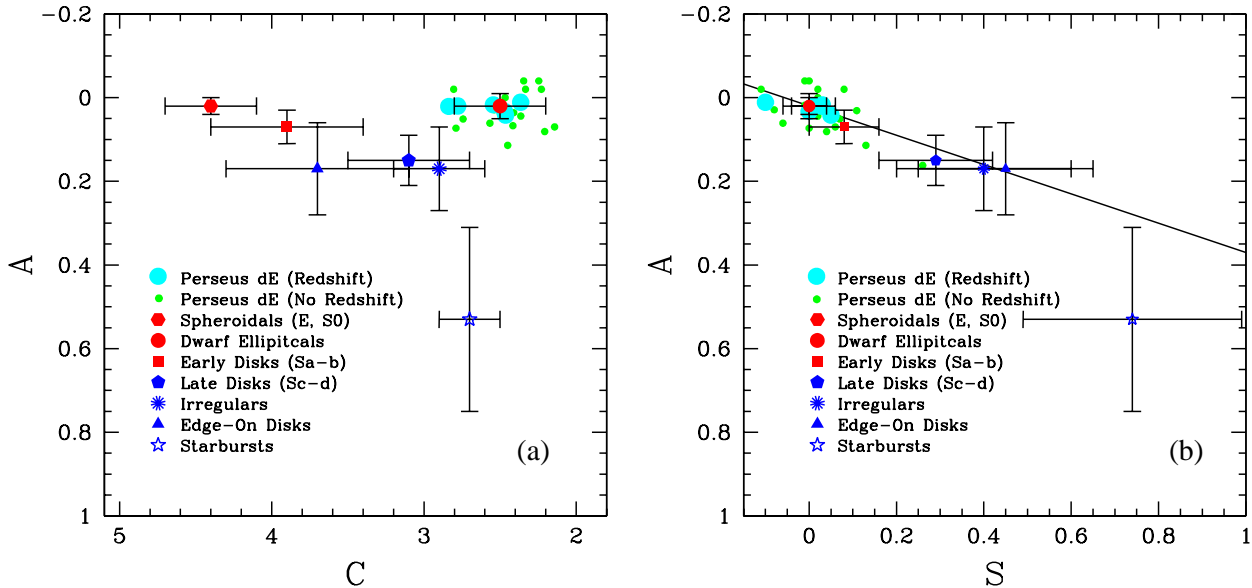}
\caption{Plot of asymmetry (A) versus concentration (C) and clumpiness (S) indices  for the Perseus dwarfs. Average C, A and S indices from \citet{cons03b} for different galaxy types are also plotted.}
\end{figure*}

The Perseus dEs in this study are found to have an average concentration value C $ = 2.6 \pm 0.6$, consistent with the average C $ = 2.5 \pm 0.3$ from \citet{cons03b} for ground based observations of dwarfs in Perseus using the WIYN 3.5m telescope. As expected for a sample of dwarf elliptical galaxies, the average A and S indices are very low (A $ = 0.03 \pm 0.04$, S $ = 0.02 \pm 0.09$), showing that these galaxies have a smooth, symmetric light distribution, lacking in internal features. Also, these dwarfs are all likely cluster members, as background galaxies often show evidence for star formation that would lead to higher S values. The CAS values for these dwarfs are plotted in Figure 2, revealing that all dwarfs in this sample have similar structures.

For comparison, we measure the values of C, A, and S for a selection of background galaxies, with the sample including both spirals and ellipticals. The galaxies in this background selection are those identified as members of the Perseus Cluster in \citet{cons03} from ground based imaging with the WIYN 3.5m telescope, based on their colours, shape, and central surface brightness. These galaxies were later rejected as cluster members by their spectroscopic redshifts \citep{me08}, or by their morphology in the $HST$ ACS imaging presented here. The average CAS values for the galaxies in the background sample are C $ = 3.0 \pm 0.7$, A $ = 0.17 \pm 0.11$, and S $ = 0.08 \pm 0.06$, significantly different to those found in the dEs.

\subsection{Surface Photometry}

We use our F814W band images to extract surface brightness, position angle, and ellipticity profiles for all the dwarfs in our sample. These were obtained using our own software, where the code fits an ellipse through a set of positions where a given surface brightness level is reached. For each surface brightness profile $I(R)$, a S\'{e}rsic brightness profile of the form:

\begin{equation}
 \mu_{I}(r) = \mu_{I,0} + 2.5b_{n}(r/R_{e})^{1/n}
\end{equation}

\noindent is fit, where $\mu_{I,0}$ is the central surface brightness, $R_{e}$ is the half light radius, $b_{n} \approx 0.848n - 0.142$, $n$ is the S\'{e}rsic index, and radius $r = \sqrt{a \times b}$, where $a$ and $b$ are the major and minor axis distances respectively. The detailed surface photometry, position angle and ellipticity profiles for two of the dwarfs in our sample are shown in Figure 3. The galaxies have central brightness peaks that cannot be fit with a S\'{e}rsic profile, so the inner 0.2'' of each galaxy are excluded from the fit. The S\'{e}rsic index $n$ for both of these dwarfs are included in Fig. 3, and are very low ($<$ 1), showing that these dwarfs have near exponential light profiles, typical for dwarf ellipticals.

We also present the ellipticities, $\epsilon = [1-(b/a)]$, for each isophote as a function of radius (Fig. 3). In general the ellipticities are constant with radius, with an average ellipticity for this sample of $<\epsilon> = 0.18 \pm 0.10$, showing that these galaxies are round systems. The position angles for the dwarfs in this sample are also constant with radius outside of the nuclei, showing that isophotal twists are not present in these dwarfs. Full analysis of the surface photometry and scaling relations of the dwarfs in this sample will be provided by De Rijcke et al., in preparation.
  
\begin{figure*}
\includegraphics[width=170mm]{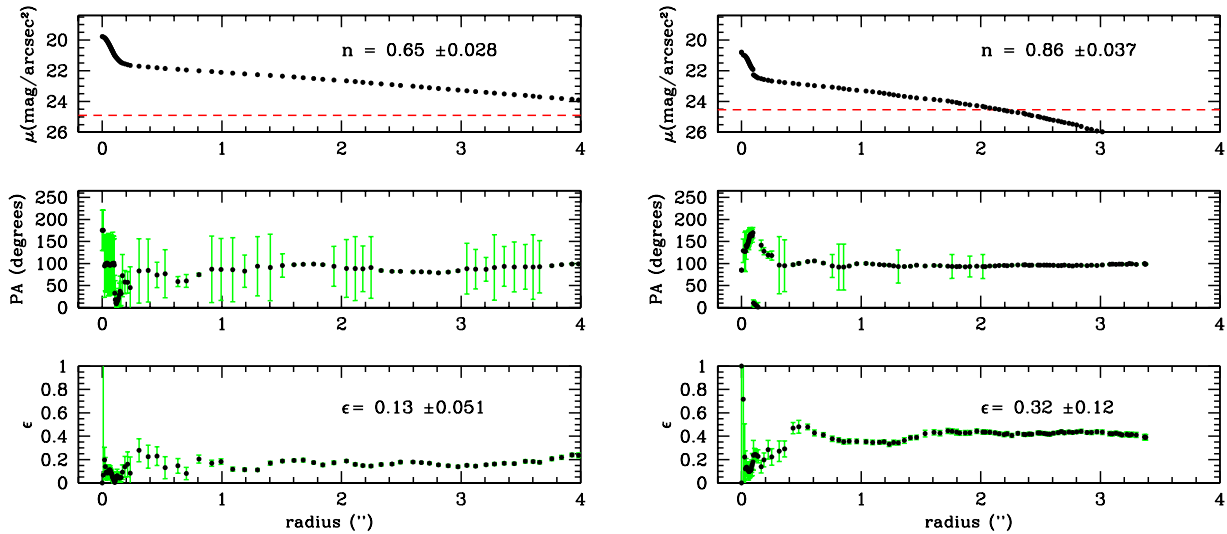}
\caption{\textit{Top to bottom}: Surface brightness, position angle and ellipticity profiles for two dEs in our sample. The S\'{e}rsic profile parameter $n$ and the ellipticity $\epsilon$ are shown for each galaxy. The dashed red line is the 1$\sigma$ error of the sky.}
\end{figure*}

We conclude from $\S$ 3.2 and 3.3, as well as Figs. 1-3, that dEs in the core of the Perseus Cluster are very smooth and symmetric, suggesting the presence of dark matter to prevent tidal disruption.

\section{Determination of dark matter content}

Based on the smoothness of our sample of dEs, we can infer that these objects might have a large dark matter component to prevent tidal disruption by the cluster potential. However, the large distance to the Perseus Cluster prevents us from easily measuring an internal velocity dispersion for these galaxies via spectroscopy. Therefore, we derive a new method for determining the dark matter content of these dwarfs by finding the minimum mass the dwarfs must have in order to prevent tidal disruption by the cluster potential.

To determine the dark matter content of our dwarfs, we assume that the dEs in this study are spherical, with no substructure, and that Newtonian gravity holds in such galaxies. With these assumptions the main properties needed to measure the dark matter content of  dEs are their sizes and their distance from the centre of the Perseus cluster, which we assume to be at the coordinates of NGC 1275 ($\alpha$ = 03:19:48.1, $\delta$ = $+$41:30:42). We take the radius of the dwarfs to be their Petrosian radii, unless the dwarf is near a bright star or galaxy, which are excluded from the study, leaving us with a sample of 25 dwarf ellipticals. However, the dark matter halos of the dwarfs likely extend beyond these Petrosian radii, placing limit on the total masses of these galaxies. We also make the approximation that the cluster itself is spherically symmetric.

Many clusters of galaxies contain hot gas that emits in X-ray wavelengths. By mapping this gas, it is possible to trace the gravitational potential and mass distribution in clusters. This allows us to estimate the cluster mass interior to each dwarf, which is also needed to estimate the mass of each dE.

\subsection{Cluster Position}

We measure the dark matter content of the dwarfs in our sample at their current cluster positions, which we find by assuming that all the dwarfs lie at the same distance as NGC 1275. We compare their coordinates to those of NGC 1275, allowing us to find their projected distance from the cluster centre. The dwarfs in our sample are not likely on circular orbits (see below), so unless we are now seeing them at the pericenters of their orbits, they will enter, at some time during their orbits, the very central regions of the cluster where the tidal forces will be even greater. The masses we measure are therefore lower limits.

We infer from their velocity dispersions that dwarfs in the core of clusters are on radial or highly elliptical orbits. One reason for this is that dwarfs in the Perseus core have a velocity dispersion of 1818~km~s$^{-1}$ \citep{me08}, higher than the value of 1324~km~s$^{-1}$ found by \citet{Stublerood99} for the cluster ellipticals (see also \citealt{conselice01}). This result is explained if we consider the angular velocities of the dwarfs when we view their orbits. If a radial orbit is viewed edge on, with the orbit pericentre closest to the observer, then the velocity dispersion observed will be highest as the dwarf will spend longer moving towards or away from the observer than on a circular orbit. The high velocity dispersion of the dwarf population indicates that they are on radial or highly elliptical orbits, and would spend part of their orbits in the innermost regions of the cluster, where they would be subject to disruptive tidal forces. If the dwarfs were on circular orbits, we would expect this velocity dispersion to be much lower. To survive these tidal forces, the dwarfs must have sufficiently high masses to prevent their disruption.

If the dwarfs are to remain intact, they must be sufficiently massive to withstand these tidal forces. At its  pericentre, a dwarf will be subject to the largest tidal forces it will experience during its orbit, and therefore this will be the point at which its mass will need to be highest to prevent tidal disruption. The orbit pericentre is where a dwarf would require the maximum mass-to-light ratio, therefore we need to estimate the closest distance from the cluster centre at which a dwarf can survive in the Perseus cluster.

Our $HST$ imaging does not cover the cluster centered galaxy NGC 1275, thus when finding the closest distance from the cluster centre at which a dwarf can survive, we include dwarfs from the \citet{cons03} study of dwarfs in the Perseus Cluster, excluding those rejected as cluster members in \citet{me08}. We also remove from the \citet{cons03} sample any galaxies which show internal structure in our $HST$ ACS imaging that confirms them as background galaxies. We measure the cluster centre distances for all remaining galaxies, and find that no dwarfs down to M$_{\rm{B}} = -12.5$ exist within a 35 kpc radius of NGC 1275. For comparison, the closest dwarf to NGC 1275 in the $HST$ ACS imaging presented here is located at a projected distance of 37.5 $\pm$ 2.6 kpc.  Therefore we take 35 kpc to be the pericentric, and thus minimum, distance from the cluster centre at which a dwarf can survive the cluster potential without being tidally disrupted. It is unlikely that all the dwarfs in our sample share the same pericentric distance, but calculating the masses the dwarfs require  at this distance gives us the maximum mass-to-light ratios we would expect to find for such galaxies in the Perseus Cluster.

\subsection{Cluster Mass}

To measure the total masses of the dwarfs in our sample, we need to know the mass of the cluster interior to each dwarf. \citet*{Mathews06} provide a model of the acceleration due to gravity, based on \textit{XMM-Newton} and $Chandra$ X-ray observations of the Perseus Cluster. The acceleration due to gravity is modeled by combining an NFW dark halo and a stellar contribution from the cluster centered galaxy NGC 1275, and fits the observed acceleration well at cluster center distances between 10 kpc and 300 kpc (See \citet{Mathews06} for details). The NFW dark matter halo acceleration is given by:

\begin{equation}
 g_{NFW} = \frac{GM_{vir}}{R^{2}}\frac{{\rm{ln}}(1+y)-y/(1+y)}{{\rm{ln}}(1+c)-c/(1+c)},
\end{equation}

\noindent where $y=cR/r_{vir}$, $c$ is the concentration, and $R$ is the distance in kpc.
 
The cluster centered galaxy NGC 1275 has a significant contribution to the mass of the cluster at small cluster centre distances, so it is necessary to include a fit to the stellar acceleration from this galaxy in the model of the acceleration due to gravity in the Perseus Cluster. The stellar acceleration $g_{*}$ = $GM_{*}(R)/R^{2}$ for the de Vaucouleurs profile presented in Figure 1 of \citet{Mathews06} is given by (in CGS units):

\begin{equation}
 g_{*} = \left[\left(\frac{R^{0.5975}}{3.206\times10^{-7}}\right)^{s}+\left(\frac{R^{0.1.849}}{1.861\times10^{-6}}\right)^{s}\right]^{1/s}
\end{equation}

\noindent with $s=0.9$ and $R$ in kpc. The total two component acceleration out to a distance $R$ from the cluster centre is given by $g(R)=g_{NFW}+g_{*}$ \citep{Mathews06}. This acceleration can be converted to an effective mass enclosed within a distance $R$, by  $M_{cl}=g(R)R^{2}/G$ (see Fig. 4). Once the mass of the cluster out to the distance of the dwarf is known, the mass a dwarf must have to survive the cluster potential can be measured. 

\subsection{Tidal Mass Measurements}

We use the above information to derive the tidal masses for our dwarfs. As a dwarf orbits the cluster centre, a star on the surface of the dwarf will become detached by the cluster tidal forces unless the dwarf has sufficient mass to prevent this occurring. We thus estimate a minimum tidal mass for the dwarf to prevent tidal disruption by the cluster potential.

Since the dwarfs are likely on radial orbits, we must take this into account when calculating their tidal masses. We use \citet{King62} to derive  the mass of a dwarf galaxy on a radial orbit in Perseus as:

\begin{equation}
\ M_{\rm{dwarf}} > \frac{r_{d}^{3}M_{cl}(R)(3+e)}{R^3}
\end{equation}

\noindent where $M_{\rm{dwarf}}$ is the total mass of the dE, $r_{d}$ is the Petrosian radius of the dwarf, $M_{cl}$(R) is the mass of the cluster interior to the dwarf, $R$ is the peri-centric distance of the dwarf from the cluster centre, and $e$ is the eccentricity of the dwarf's orbit. The eccentricity $e$ is defined as:

\begin{equation}
e = 1 - \frac{2}{(R_{a}/R_{p})+1}
\end{equation}

\noindent where $R_{a}$ is the cluster centre distance at the apocentre and $R_{p}$ is the cluster centre distance at the pericentre. For $e = 0$ the orbit will be circular, and for $0 < e < 1$ the orbit will be elliptical. The exact value of $e$ is not important as we do not know the exact orbits of the dwarfs, thus we adopt a value of $e = 0.5$ as a compromise.

We calculate theoretical masses for galaxies of various radii using Eqn 5. As we do not know the tidal radii of our dwarfs, we instead measure  the masses the dwarfs would need to prevent them being disrupted at their Petrosian radii. Figure 4 shows the predicted masses of galaxies of different radii as a function of distance from the cluster centre and size.

\begin{figure}
\includegraphics[width=84mm]{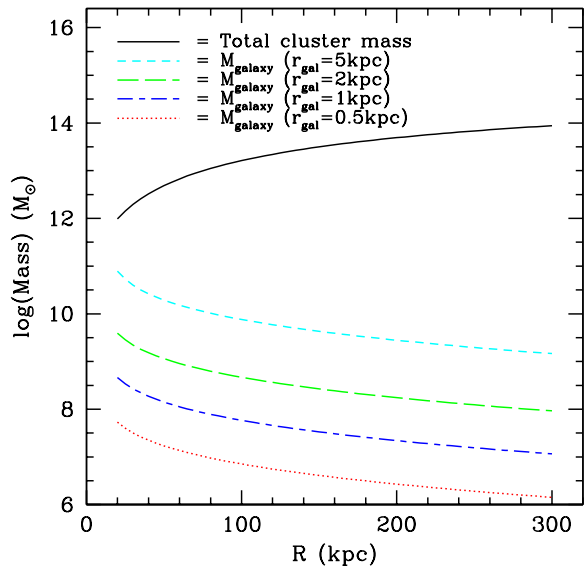}
\caption{Relationship between predicted galaxy mass and cluster centre distance for galaxies of varying radii, found using our method. The masses of galaxies with radii of 0.5, 1, 2, and 5 kpc are plotted, as well as the total cluster mass, as a function of distance from the cluster centre. The model of the cluster mass is valid for the region 10 kpc $\leq R \leq$ 300 kpc.}
\end{figure} 

In order for stellar to total mass ratios to be estimated for our dwarfs, the luminous mass of each dwarf must be measured. This is determined using the method of \citet{bell01}. First, the colour of each dwarf is measured within a central 1'' radius. The colour is then converted to a stellar mass-to-light ratio by using the mass-dependent galaxy formation epoch model with a scaled-down Saltpeter IMF from \citet{bell01}. Stellar masses are then be found for the dwarfs by converting their absolute magnitudes to solar luminosity, and then multiplying this by the stellar mass-to-light ratio.

\section{Results}

\subsection{Local Group}

We test the method presented in this paper by calculating the masses of six Milky Way dSphs (Carina, Leo I, Fornax, Sculptor, Draco, and Ursa Minor) within their tidal radii. The Milky Way is assumed to be a point mass of 10$^{12}$ M$_{\odot}$ for this calculation. We measure the dark matter content of these dwarfs by assuming they are on radial orbits, both at their current distances from the MW, and by assuming that they pass through their computed orbit pericentre. Proper motions have been found by various authors for the MW dSphs we examine here \citep{oden,piatek03,piatek05,piatek06,piatek07,sohn}, and these authors have found perigalacticon distances and orbital eccentricities which we use in our calculations. We compare our results with the masses obtained using kinematics presented in \citet{strigari07}. The results are presented in Figure 5, with errors due to uncertainties in the distances to the dSphs and their tidal radii.

\begin{figure}
\includegraphics[width=84mm]{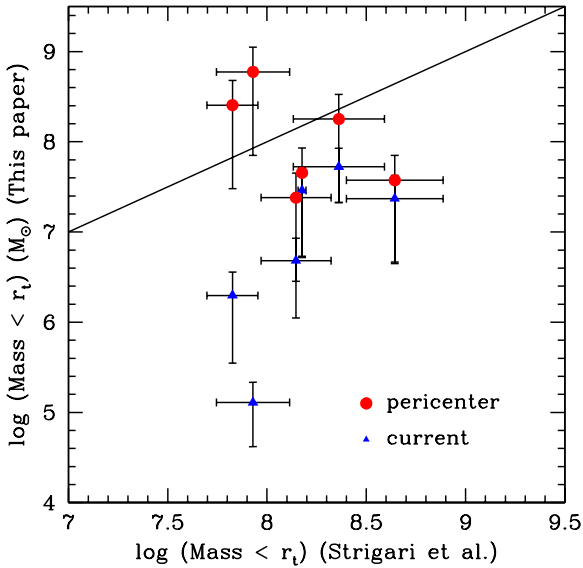}
\caption{Masses for the Milky Way dSphs using the method presented in this paper compared to the results of \citet{strigari07}, obtained using kinematics. The masses are calculated at both the current positions of the dSphs (triangles), and at their orbit pericenters (circles). The solid line is a 1:1 relation.}
\end{figure}

It is clear from Fig. 5 that the masses of the MW dSphs found at their orbit pericenters using the tidal method presented here match their dynamical masses  more closely than those obtained at their current distances. When the tidal masses for the MW dSphs are calculated at their current positions, the least massive galaxies are found to have masses up to a factor of 100 times smaller than their dynamical masses. For Leo I, Carina and Draco, this yields mass-to-light ratios insufficiently high to prevent disruption by tidal forces at their current distances from the MW. 

We also calculate the masses the dSphs require to survive the tidal potential of the Milky Way at a distance of 38 kpc. This value was chosen as the dSph Willman 1 \citep{will} is located at a distance of 38 $\pm$ 7 kpc from the Milky Way, and dwarfs interior to this radius (e.g. Sagittarius) are tidally disrupted. At this distance, there is good agreement between the masses calculated for the dSphs  using our tidal method, and the masses obtained via kinematics.  Therefore it seems that the dynamical masses for the LG dSphs reflect the masses they would require to withstand the tidal forces at the pericentre of their orbits around the Milky Way. This also verifies that our method does work and at minimum gives a lower limit to the masses of dEs.

\subsection{Tidal radii}

Before we discuss our calculated M/L ratios, we examine the expected tidal radii of our dwarfs. We estimate the tidal radii for each of our dwarfs by assuming that they only contain luminous material. If a dwarf has a smaller tidal radius than observed radius, dark matter must be present in the galaxy in order to prevent it from being tidally disrupted and dispersed by the cluster potential. We calculate the tidal radii at the projected cluster centre distances of the dwarfs, which we take to be the projected distances that the dwarfs are currently from the cluster centered galaxy NGC 1275.

When calculating the tidal radii, we assume that the dwarfs only have a mass $m_{stellar}$, and are only made of luminous material with no dark matter component. The tidal radii of the dwarfs can be calculated using Eqn. 5, with errors due to uncertainties in the luminous mass of the dwarf, the distance to the cluster, and the mass of the cluster accounted for. Figure 6 shows the predicted tidal radii of the dwarfs at their current cluster positions, plotted against their observed Petrosian radii, with a 1:1 relation plotted. The majority of the dwarfs in this study have larger observed radii than their predicted tidal radii, which suggests that these dwarfs must contain dark matter to prevent disruption at their current positions.

\begin{figure}
\includegraphics[width=84mm]{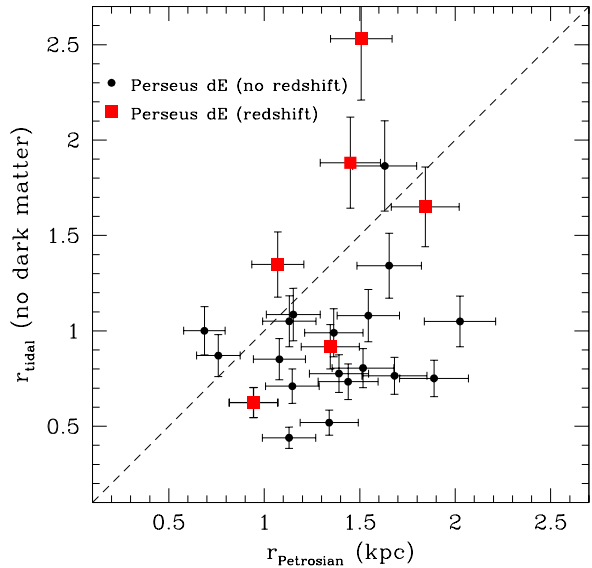}
\caption{Relationship between the predicted tidal and Petrosian radii. The dashed line is a 1:1 relation. Dwarfs below this line have larger observed radii than their expected tidal radii, and would therefore need dark matter to prevent tidal disruption. }
\end{figure}

\subsection{Mass-to-light ratios}

There exists a relationship for Local Group dwarf spheroidals such that fainter dwarfs have higher M/L ratios. The faintest dwarfs with M$_{\rm{V}} < -4$ have calculated mass-to-light ratios approaching 1000 M$_{\odot}$/L$_{\odot,V}$ \citep{simon07}. Using the dE sample presented in this paper, we investigate the relationship between M/L ratio and luminosity in a dense cluster environment. We measure the mass-to-light ratios for the dwarfs in our sample using the new method presented in this paper, with the results shown in Figure 7. Also included on this plot for comparison are M/L ratios for the Milky Way dSphs,  obtained using kinematics taken from \citet{strigari07}. We measure the mass-to-light ratios for the Perseus dwarfs at their current projected cluster centre distances. We also consider the masses the dwarfs must have to prevent disruption in the very centre of the cluster, at a distance of 35 kpc from NGC 1275. We use an eccentricity $e = 0.5$. 

\begin{figure*}
\includegraphics[width=168mm]{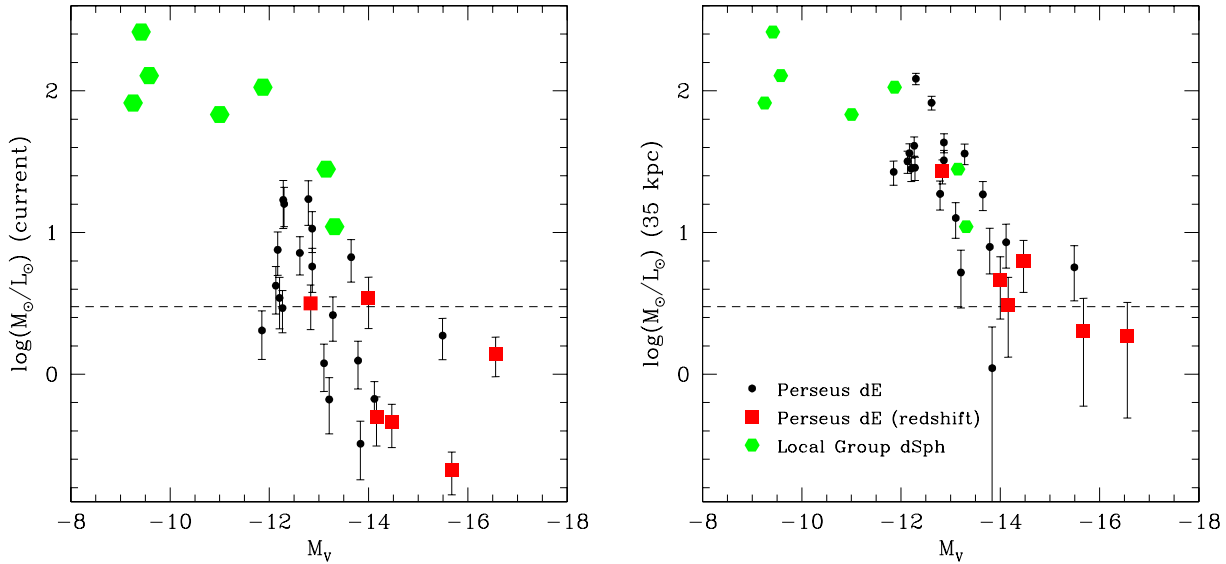}
\caption{Relationship between mass-to-light ratios and $M_{V}$ for Perseus dEs and Milky Way dSphs. The left hand plot shows M/L for the dwarfs at their current cluster position, and the right hand plot shows M/L for the dwarfs at a distance of 35 kpc from NGC 1275. Values of M/L for the Milky Way dSphs are taken from \citet{strigari07}, and were obtained using kinematics. The dashed line in each plot is at a M/L of 3, which we take, within our errors, to be the mass-to-light ratio at which no dark matter is required to prevent the dwarf being tidally disrupted.}
\end{figure*}

Figure 7 shows that at their current cluster positions, twelve of the dwarfs in our sample require dark matter, with the remaining dwarfs having mass-to-light ratios smaller than 3, indicating they do not require dark matter within our errors at their current distances from the cluster centre. When we measure the maximum  mass-to-light ratios the dwarfs could have, i.e. the mass they require to prevent disruption at a distance of 35 kpc from the cluster centre, we find that all but three require dark matter (Figure 7b).

 There is a clear correlation between mass-to-light ratio and the luminosity of the dwarfs, such that the faintest dwarfs require the largest fractions of dark matter to remain bound. This is to expected, as the fainter a galaxy is, the fewer stars (and therefore luminous mass) it will contain. This result is also seen for the Milky Way dSphs, with both cluster and MW dwarfs requiring similar amounts of dark matter at a given luminosity .

 We further calculate the distance from the cluster centre at which the dwarfs do not require dark matter. The tidal masses of the dwarfs are calculated for distances between $R = 5$ kpc and $R = 500$ kpc from the cluster centre, and the distance at which M$_{\odot}$/L$_{\odot} =3$ is recorded. The value M$_{\odot}$/L$_{\odot} =3$ is chosen to take into account our errors caused by uncertainties in the distance to the cluster and the stellar mass of the dwarfs. These results are plotted in Figure 8. 

\begin{figure}
\includegraphics[width=84mm]{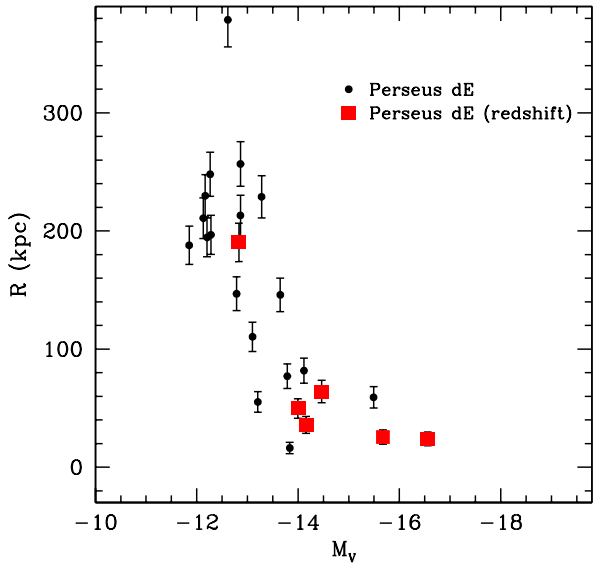}
\caption{Distance from the cluster centre at which M$_{\odot}$/L$_{\odot}$ falls below 3 plotted against $M_{\rm{V}}$ for the dwarfs in our sample. If these dwarfs are at smaller distances from the cluster centre than those plotted above, they would require dark matter}
\end{figure}

Fig. 8 shows that the fainter the dwarf, the further from the cluster centre it has to be for its stellar mass alone to be sufficient to prevent disruption by the cluster potential. Errors are due to uncertainty in the distance to the Perseus Cluster and therefore the sizes of the dEs. The brightest dwarfs in our sample have a massive enough stellar population that they do not require dark matter to survive at any position in the cluster.

\section{Discussion}

All the dwarfs presented here are remarkably smooth in appearance, and this smoothness can be used to partially explain their origin.  It is thought that some dwarf galaxies in clusters are the remnants of stripped disks or dwarf irregulars. Through harassment and interactions, these progenitors become morphologically transformed into dEs \citep{moore}. If dwarfs form in this way, then it is likely that internal structure would remain in the galaxy (e.g. \citealt{sven03}). The dwarfs in our sample show no such evidence, with average CAS values of  $<\rm{A}>~=~0.03~\pm~0.04$, and $<\rm{S}>~=~0.02~\pm~0.09$. Therefore it is likely that these dwarfs form by some other method, although more than one formation scenario may be required as clusters are known to contain dwarfs with a range of ages and metallicities (e.g. \citealt{heldmould,rakos01,cons03,dolf07,me08}).

We find that at their current projected distances from the cluster centre, approximately 1/2 of the dwarfs in our sample require dark matter to prevent tidal disruption  by the cluster potential, and this includes nearly all of the faint systems (M$_{V} > 13$). Dwarfs likely reside in dark matter halos that extend well beyond their stellar component, therefore the total dark matter in the cluster dwarfs, and their mass-to-light ratios, will be further increased if this dark matter is taken into account. It is clear, however, that dwarf ellipticals in all environments need large quantities of dark matter to survive. Dwarfs in every environment follow the same scaling relations (De Rijcke et al. 2008, in preparation), suggesting that dwarfs are immune to environmental influences, possibly due to their large M/L ratios.

The faintest cluster dwarfs are found to have total mass-to-light ratios similar to those in the the Local Group.  The results presented in this paper for the cluster dwarfs place a lower limit on the mass-to-light ratios for the dwarfs, as we do not consider any mass that may exist outside their Petrosian radii. Therefore we can infer that dwarf galaxies in dense cluster environments need similar mass-to-light ratios as the Local Group dwarfs to remain bound.  

In groups of galaxies, dwarfs are most commonly found as satellites of larger galaxies (e.g. \citealt{bin90,mat98}), and their close proximity to large host systems means these satellite galaxies need dark matter to prevent tidal disruption by their host galaxies. In clusters, dwarfs are not distributed around the giant galaxies, but follow the general potential. The dense cluster environment would have a destructive effect on dwarf ellipticals if they did not contain large dark matter components to prevent tidal disruption. 

Several methods can disrupt dwarfs in the centers of clusters. One such process is harassment \citep*{moore}. High speed interactions with other galaxies, such as the giant ellipticals present in the Perseus Cluster (e.g. NGC 1275, NGC 1272), would impart energy to the dwarf during interaction and increase its internal energy. Such interactions are unlikely to destroy a dwarf unless it passes within $\sim 10$ kpc of a giant elliptical \citep{conselice01}, but would cause tidal disruption to the dwarf if it were not massive enough to withstand the increase in its internal energy. 

We find that the mass-to-light ratios of dEs in the centre of the Perseus cluster increase as the luminosity of the galaxies decrease, such that the least luminous dwarfs need the most dark matter (M/L $\sim 15$ at M$_{\rm{V}} = -12.5$) to prevent tidal disruption by the cluster potential. This is clearly seen in the Local Group, with the mass-to-light ratios for the brighter Local Group dEs lower than those of the LG dSphs, with values of M$_{\odot}$/L$_{\odot}$ $\approx 3-4$ for NGC 147, 185 and 205 \citep{derijcke06}. One prediction of Cold Dark Matter theory is that the least massive halos are denser than more massive systems \citep{NFW}. The least luminous dwarfs  would have the most dense dark matter halos, and therefore highest mass-to-light ratios. 

Determining the dark matter content of dwarfs using the method presented here has several advantages over using kinematics. Measuring the dark matter content of a  dwarf elliptical using kinematics makes the assumption that the galaxy is in virial equilibrium, so the overall mass of the system can be determined from the observed velocity dispersion. However, if the galaxy is perturbed by tides, the kinematics of its stellar population will be disturbed and the assumption that it is in virial equilibrium will be incorrect. The presence of unresolved tidal tails may also contaminate the data \citep{Lokas2008}. It is also not easy to obtain spectra in the fainter outer regions of dwarf ellipticals, so their mass-to-light ratios can only be estimated for the central regions of the galaxy. The method presented here for determining the dark matter content of cluster dwarfs without using kinematics avoids these problems. 

Modified Newtonian Dynamics (MOND: \citealt{MOND}) has been used to argue that the difference between the kinematic mass and the baryonic mass of galaxies is due to the Newtonian law of gravity not governing the dynamics of such systems. We have shown here, without the use of kinematics, that dwarf galaxies in clusters require the presence of dark matter to survive in the dense cluster environment. Therefore it seems likely that dark matter is a real component of dwarf elliptical galaxies in all environments. 

\section{Summary}

We have identified from $HST$ ACS imaging a total of 29 dwarf elliptical galaxies in the core of the Perseus Cluster of galaxies, 17 of which are newly discovered. All of these galaxies are remarkably smooth, symmetrical,  and round in appearance. To quantify the degree of their smoothness, we measure the asymmetry and clumpiness parameters for these galaxies, as well as parametric fitting presented in $\S$3.3. Our sample of dwarfs show little to no evidence for internal structure, with A $= 0.03 \pm 0.04$ and S = $0.02 \pm 0.09$. The surface brightness profiles for the dwarfs show that they are round in shape with $\epsilon_{av} = 0.18 \pm 0.10$, with position angles that stay constant with radius, again highlighting the smoothness and lack of features in these dwarfs. 

The fact that these dwarfs are not morphologically disturbed suggests that they must contain a large proportion of dark matter to prevent tidal disruption, and we derive a new method for measuring the dark matter content of these objects based purely on their smoothness and radius. We show that in order to remain bound systems whilst at small cluster centre distances, faint dwarf ellipticals in a cluster environment must have a substantial proportion of dark matter to prevent tidal disruption/distortion, with (M/L) ranging between 1 for the brightest dwarfs in our sample up to 120 for the faintest galaxies.

\section*{Acknowledgments}

SJP acknowledges the support of a STFC studentship. SDR is a Postdoctoral Fellow of the Fund for Scientific Research- Flanders (Belgium)(F.W.O.). This research has made use of the NASA/IPAC Extragalactic Database (NED) which is operated by the Jet Propulsion Laboratory, California Institute of Technology, under contract with the National Aeronautics and Space Administration.


\begin{thebibliography}{99}
\bibitem[\protect\citeauthoryear{Bell \& de Jong}{2001}]{bell01} Bell E.F., de Jong R.S., 2001, ApJ, 550, 212
\bibitem[\protect\citeauthoryear{Bershady, Jangran \& Conselice}{Bershady et al.}{2000}]{Bershady00} Bershady M.A., Jangren A., Conselice C.J., 2000, AJ, 119, 2645
\bibitem[\protect\citeauthoryear{Binggeli, Tarenghi  \& Sandage}{1990}]{bin90} Binggeli B., Tarenghi M., Sandage A., 1990, A\&A, 228, 42 
\bibitem[\protect\citeauthoryear{Binney \& Tremaine}{2008}]{BT} Binney J., Tremaine S., 2008, Galactic Dynamics, 2nd edn. Princeton University Press, Princeton, NJ
\bibitem[\protect\citeauthoryear{Conselice, Bershady \& Jangren}{Conselice et al.}{2000}]{Conselice00} Conselice C.J., Bershady M.A., Jangren A., 2000, ApJ, 529, 886
\bibitem[\protect\citeauthoryear{Conselice, Gallagher \& Wyse}{Conselice et al.}{2001a}]{conselice01} Conselice C.J., Gallagher J.S. III, Wyse R.F.G., 2001a, ApJ, 559, 791
\bibitem[\protect\citeauthoryear{Conselice, Gallagher \& Wyse}{Conselice et al.}{2001b}]{conselice01b} Conselice C.J., Gallagher J.S. III, Wyse R.F.G., 2001b, AJ, 122, 2281
\bibitem[\protect\citeauthoryear{Conselice, Gallagher \& Wyse}{Conselice et al.}{2002}]{conselice02} Conselice C.J., Gallagher J.S. III, Wyse R.F.G., 2002, AJ, 123, 2246
\bibitem[\protect\citeauthoryear{Conselice, Gallagher \& Wyse}{Conselice et al.}{2003}]{cons03} Conselice C.J., Gallagher J.S. III, Wyse R.F.G., 2003, AJ, 125, 66
\bibitem[\protect\citeauthoryear{Conselice}{2003}]{cons03b} Conselice C.J., 2003, ApJS, 147, 1
\bibitem[\protect\citeauthoryear{De Rijcke \& Dejonghe}{2002}]{sven02} De Rijcke S., Dejonghe H., 2002, MNRAS, 329, 829
\bibitem[\protect\citeauthoryear{De Rijcke et al.}{2003}]{sven03} De Rijcke S., Dejonghe H., Zeilinger W.W., Hau G.K.T., 2003, A\&A, 400, 119
\bibitem[\protect\citeauthoryear{De Rijcke et al.}{2006}]{derijcke06} De Rijcke S., Prugniel P., Simien F., Dejonghe H., 2006, MNRAS, 369, 1321
\bibitem[\protect\citeauthoryear{Held \& Mould}{1994}]{heldmould} Held E.V., Mould J.R., 1994, AJ, 107, 1307
\bibitem[\protect\citeauthoryear{Klessen \& Kroupa}{1998}]{KlessenKroupa} Klessen R.S., Kroupa P., 1998, ApJ, 498, 143
\bibitem[\protect\citeauthoryear{Kron}{1995}]{Kron95} Kron R.G., 1995, The Deep Universe: Saas-Fee Advanced Course 23. Springer, New York
\bibitem[\protect\citeauthoryear{{\L}okas, Mamon \& Prada}{{\L}okas et al.}{2005}]{lokas05}{\L}okas E.L., Mamon G.A., Prada F., 2005, MNRAS, 363, 918
\bibitem[\protect\citeauthoryear{{\L}okas et al.}{2008}]{Lokas2008} {\L}okas E.L., Klimentowski J., Kazantzidis S., Mayer L., 2008, arXiv:0804.0204
\bibitem[\protect\citeauthoryear{King}{1962}]{King62} King I., 1962, AJ, 67, 471
\bibitem[\protect\citeauthoryear{Mateo}{1998}]{mat98} Mateo M.L., 1998, ARA\&A, 36, 435
\bibitem[\protect\citeauthoryear{Mateo, Olszewski \& Walker}{Mateo et al.}{2008}]{mateo08} Mateo M., Olszewski E.W., Walker M.G., 2008, ApJ, 675, 201
\bibitem[\protect\citeauthoryear{Mathews, Faltenbacher \& Brighenti}{Mathews et al.}{2006}]{Mathews06} Mathews W.G., Faltenbacher A., Brighenti F., 2005, ApJ, 638, 659
\bibitem[\protect\citeauthoryear{Michielsen et al.}{2007}]{dolf07} Michielsen D., et al., 2007, ApJ, 670, L101 
\bibitem[\protect\citeauthoryear{Milgrom}{1983}]{MOND} Milgrom M., 1983, ApJ, 270, 365
\bibitem[\protect\citeauthoryear{Moore, Lake \& Katz}{Moore et al.}{1998}]{moore} Moore B., Lake G., Katz N., 1998, ApJ, 495, 139
\bibitem[\protect\citeauthoryear{Navarro, Frenk \& White}{Navarro et al.}{1997}]{NFW} Navarro J.F., Frenk C.S., White S.D.M., 1997, ApJ, 490, 493
\bibitem[\protect\citeauthoryear{Odenkirchen et al.}{2001}]{oden} Odenkirchen M., et al., 2001, AJ, 122, 2538
\bibitem[\protect\citeauthoryear{Penny \& Conselice}{2008}]{me08} Penny S.J., Conselice C.J., 2008, MNRAS, 383, 247
\bibitem[\protect\citeauthoryear{Petrosian}{1976}]{Pet76} Petrosian V., 1976, ApJ, 209, L1
\bibitem[\protect\citeauthoryear{Piatek et al.}{2003}]{piatek03} Piatek S., Pryor C., Bristow P., Olszewski E.W., Harris H.C., Mateo M., Minniti D., Tinney C.G., 2003, AJ, 126, 2346
\bibitem[\protect\citeauthoryear{Piatek et al.}{2005}]{piatek05} Piatek S., Pryor C., Bristow P., Olszewski E.W., Harris H.C., Mateo M., Minniti D., Tinney C.G., 2005, AJ, 130, 95
\bibitem[\protect\citeauthoryear{Piatek et al.}{2006}]{piatek06} Piatek S., Pryor C., Bristow P., Olszewski E.W., Harris H.C., Mateo M., Minniti D., Tinney C.G., 2006, AJ, 131, 1445
\bibitem[\protect\citeauthoryear{Piatek et al.}{2007}]{piatek07} Piatek S., Pryor C., Bristow P., Olszewski E.W., Harris H.C., Mateo M., Minniti D., Tinney C.G., 2007, AJ, 133, 844
\bibitem[\protect\citeauthoryear{Rakos et al.}{2001}]{rakos01} Rakos K., Schombert J., Maitzen H.M., Prugovecki S., Odell A., 2001, AJ, 121, 1974
\bibitem[\protect\citeauthoryear{Schelgel et al.}{1998}]{Schlegel} Schlegel D.J., Finkbeiner D.P., Davis M., 1998, ApJ, 500, 525
\bibitem[\protect\citeauthoryear{S\'{e}gall et al.}{2007}]{segall07} S\'{e}gall M., Ibata R.A., Irwin M.J., Martin N.F., Chapman S., 2007, MNRAS, 375, 831
\bibitem[\protect\citeauthoryear{Simon \& Geha}{2007}]{simon07} Simon J.D., Geha M., 2007, ApJ, 670, 313
\bibitem[\protect\citeauthoryear{Sirianni et al.}{2005}]{sirianni} Sirianni M., et al., 2005, PASP, 117, 1049
\bibitem[\protect\citeauthoryear{Sohn et al.}{2007}]{sohn} Sohn, S.T., et al., 2007, ApJ, 663, 960
\bibitem[\protect\citeauthoryear{Strigari et al.}{2007}]{strigari07} Strigari L.E., Bullock J.S., Kaplinghat M., Diemand J., Kuhlen M., Madau P., 2007, ApJ, 669, 676
\bibitem[\protect\citeauthoryear{Struble \& Rood}{1999}]{Stublerood99} Struble M.F., Rood H.J., 1999, ApJS, 125, 36
\bibitem[\protect\citeauthoryear{Willman et al.}{2006}]{will} Willman B., et al., 2006, astro-ph/0603486
\end{thebibliography}
\end{document}